\documentclass[12pt]{article}

\usepackage{amsthm}
\usepackage{amssymb}
\usepackage{natbib}
\usepackage{graphicx}
\usepackage{url}
\usepackage{algorithm2e}
\usepackage{mathtools}
\usepackage{times}
\usepackage{adjustbox}

\usepackage{mathtools}
\usepackage{times}
\usepackage{adjustbox}

\newtheorem{assumption}{Assumption}
\newtheorem{theorem}{Theorem}
\newcommand{\E}{\mathbb{E}}
\newcommand{\1}{\mathbf{1}}
\renewcommand{\P}{\mathbb{P}}

\newcommand{\Var}{\mathrm{Var}}

\usepackage{abstract}

\usepackage{titlesec}
\titleformat*{\section}{\large\bfseries}
\titleformat*{\subsection}{\normalsize\bfseries}

\usepackage{xcolor}
\definecolor{skylight}{RGB}{45, 90, 160}

\usepackage[colorlinks=true, linkcolor=skylight, citecolor=skylight, urlcolor=black]{hyperref}

\usepackage{mathtools}
\usepackage{times}
\usepackage{adjustbox}

\usepackage[T1]{fontenc}
\usepackage[utf8]{inputenc}
\usepackage{newtxtext, newtxmath}
\usepackage[protrusion=true, expansion=true]{microtype}
\usepackage{setspace}

\usepackage[english]{babel}

\usepackage{natbib}
\setcitestyle{round,authoryear,semicolon}

\makeatletter
\renewcommand{\@fnsymbol}[1]{\ifcase#1\or a\or b\or c\or d\or e\else\@ctrerr\fi}
\makeatother

\usepackage[margin=1in]{geometry}

\usepackage{booktabs}
\usepackage{multirow}
\usepackage[hang, font= small, labelfont=bf, textfont=normalfont, position=bottom]{caption}
\usepackage{enumerate}
\usepackage{graphicx}
\usepackage{longtable}
\usepackage{threeparttable}

\usepackage{abstract}

\usepackage{titlesec}
\titleformat*{\section}{\large\bfseries}
\titleformat*{\subsection}{\normalsize\bfseries}
\title{
{\Large\bfseries{
Automatic debiased machine learning and sensitivity analysis for sample selection models}
}}

\author{
\normalsize
Jakob Bjelac
\thanks{
Putlitzstraße 2, 10551 Berlin, c/o Valdez, Germany; e-mail: \texttt{\href{mailto:jakob.bjelac@outlook.com}{jakob.bjelac@outlook.com}}
}
\and
\normalsize
Victor Chernozhukov
\thanks{
Department of Economics, Massachusetts Institute of Technology, 50 Memorial Drive, Cambridge, MA 02142, USA; e-mail: \texttt{\href{mailto:vchern@mit.edu}{vchern@mit.edu}}
}
\and
\normalsize
Phil-Adrian Klotz
\thanks{
Düsseldorf Institute for Competition Economics, Heinrich Heine University Düsseldorf, Universitätsstr. 1, 40225 Düsseldorf, North Rhine--Westphalia, Germany; e-mail: \texttt{\href{mailto:klotz@dice.hhu.de}{klotz@dice.hhu.de}}
}
\and
\normalsize
Jannis Kueck
\thanks{
Düsseldorf Institute for Competition Economics, Heinrich Heine University Düsseldorf, Universitätsstr. 1, 40225 Düsseldorf, North Rhine--Westphalia, Germany; e-mail: \texttt{\href{mailto:kueck@dice.hhu.de}{kueck@dice.hhu.de}}
}
\and
\normalsize
Theresa M. A. Schmitz
\thanks{Chair of Statistics and Econometrics, Heinrich Heine University Düsseldorf, Universitätsstr. 1, 40225 Düsseldorf, North Rhine--Westphalia, Germany; e-mail: \texttt{\href{mailto:theresa.schmitz@hhu.de}{theresa.schmitz@hhu.de}}}
}

\date{\small\today}

\begin{document}

\maketitle

\thispagestyle{empty}

\renewcommand{\abstractname}{\vspace{-4.03 em}}

\begin{abstract}
\noindent
In this paper, we extend the Riesz representation framework to causal inference under sample selection, where both treatment assignment and outcome observability are non-random. 
Formulating the problem in terms of a Riesz representer enables stable estimation and a transparent decomposition of omitted variable bias into three interpretable components: a data-identified scale factor, outcome confounding strength, and selection confounding strength.
For estimation, we employ the ForestRiesz estimator, which accounts for selective outcome observability while avoiding the instability associated with direct propensity score inversion.
We assess finite-sample performance through a simulation study and show that conventional double machine learning approaches can be highly sensitive to tuning parameters due to their reliance on inverse probability weighting, whereas the ForestRiesz estimator delivers more stable performance by leveraging automatic debiased machine learning.
In an empirical application to the gender wage gap in the U.\,S., we find that our ForestRiesz approach yields larger treatment effect estimates than a standard double machine learning approach, suggesting that ignoring sample selection leads to an underestimation of the gender wage gap. Sensitivity analysis indicates that implausibly strong unobserved confounding would be required to overturn our results. Overall, our approach provides a unified, robust, and computationally attractive framework for causal inference under sample selection.

\footnotesize
\vspace{1.06 em}
\end{abstract}

\clearpage

\newpage

\section{Introduction}

In many empirical studies, researchers face the challenge that outcomes are only observed for a subset of the sample population. Returns to education studies observe wages only for employed individuals. Job training evaluations miss earnings data for unemployed participants. Clinical trials lose patients before outcome measurement and also educational interventions suffer when students do not take standardized tests. This phenomenon, commonly referred to as sample selection or outcome attrition, complicates the estimation of causal effects \citep{heckman1976common,heckman1979sample,hausman1979attrition,little1995modeling}.
The problem becomes even more complex when treatment assignment is itself non-random. In such cases, researchers confront what \citet{Bia2024} describe as the “double selection problem”, involving both selection into treatment and selection into outcome observability. 
Standard methods for confounding adjustment, such as regression or propensity score weighting, fail when outcomes are selectively missing. Even inverse probability weighting, which addresses treatment selection, requires modification to handle missing outcomes \citep{robins1994estimation,hernan2004structural}.
The machine learning literature offers powerful tools for high-dimensional covariate adjustment, yet it also introduces new challenges. In particular, regularization inherent in machine learning estimators can induce bias that invalidates standard inference procedures unless appropriate orthogonality conditions are imposed \citep{chernozhukov2018double}. \citet{Bia2024} address this issue by deriving a Neyman-orthogonal score function for treatment effect estimation in the presence of sample selection. \citet{dolgikh2025double} also propose double machine learning estimators for treatment effect estimation in the multivariate sample selection model with ordinal selection equations. 
An alternative approach is provided by the Riesz representation theorem. Instead of relying on Neyman-orthogonal score functions, target parameters can be characterized through unique weighting functions called Riesz representers \citep{chernozhukov2022debiased}. The Riesz framework offers several advantages: it avoids unstable propensity score inversions, enables direct estimation via variational or adversarial methods, and naturally accommodates sensitivity analysis. Existing applications include the estimation of the Average Treatment Effect (ATE) and other policy-relevant causal parameters in settings without selection \citep{chernozhukov2022debiased} and the estimation of the Average Treatment Effect on the Treated (ATT) in Difference-in-Differences models \citep{bach2025sensitivityanalysistreatmenteffects}.

This paper extends the Riesz representation methods to sample selection models. We show that the ATE identified by \citet{Bia2024} via efficient scores also admits identification through a Riesz representation. The corresponding representer takes the form of inverse probability weights that adjust simultaneously for treatment assignment and sample selection. 
The Riesz representer framework is particularly useful for analyzing bias induced by unobserved selection confounding. It yields an interpretable decomposition of the omitted variable bias. Building on  \citet{cinelli2020making} and \citet{chernozhukov2022long}, we express the bias as the product of three terms: (i) a scale factor identifiable from observed data, (ii) the strength of confounding in the outcome equation, and (iii) the strength of confounding in the selection equation. This decomposition delivers sharp bounds on the magnitude of bias without requiring the specification of the full joint distribution of unobservables. 
A key insight is that observed covariates provide natural benchmarks for calibrating these sensitivity parameters \citep{imbens2003sensitivity,altonji2005selection,oster2019unobservable}. 
In our simulation study, we investigate the finite-sample behavior of the proposed ForestRiesz estimator and find that it performs well in finite samples when estimating the ATE.
As an empirical contribution, we study the gender wage gap in the U.\,S. using data from the American Community Survey. We find that our ForestRiesz approach yields larger treatment effect estimates than a standard double machine learning approach which does not account for sample selection. This suggests that ignoring sample selection leads to an underestimation of the gender wage gap, as wage reporting behavior differs systematically between female and male respondents.

\section{Identification under Confounding and Sample Selection}\label{sampleselectionmodel}

Estimation of treatment effects is fundamental to empirical research in economics, medicine, and the social sciences. This section introduces the Average Treatment Effect (ATE) within the potential outcomes framework and examines the problem of sample selection, which occurs when outcome data are missing for some units in the analysis.

\subsection{Defining Causal Effects: The Potential Outcomes Framework}

To formally define causal effects, we rely on the Potential Outcomes framework, often associated with \citet{rubin1974estimating, rubin1977assignment}. Let $D$ be a variable that represents the treatment status assigned to an individual unit $i$. For clarity, we consider a binary treatment where $D_i=1$ if unit $i$ receives the treatment and $D_i=0$ if unit $i$ receives the control, though the framework readily extends to multiple discrete treatments $d \in \{0, 1, ..., Q\}$.
For each unit \(i\), we define two potential outcomes: \(Y_i(1)\) is the outcome that unit \(i\) would have experienced under treatment (\(D_i = 1\)), while \(Y_i(0)\) is the outcome it would have experienced under control (\(D_i = 0\)).
We assume SUTVA: unit $i$'s potential outcomes are unaffected by other units' treatment assignments, and $Y_i(1)$ and $Y_i(0)$ are well-defined for each unit \citep{rubin1980randomization}.
Given the impossibility of observing individual treatment effects directly for a single unit, empirical research typically focuses on estimating average causal effects across a population or subpopulation. The most common target parameter is the Average Treatment Effect (ATE) for the entire population:
\[
\text{ATE} = \mathbb{E}[Y(1) - Y(0)] = \mathbb{E}[Y(1)] - \mathbb{E}[Y(0)],
\]
where the expectation $\mathbb{E}[\cdot]$ is taken over the distribution of units in the population of interest. 

\subsection{The Sample Selection Problem}

In many practical applications, the outcome variable $Y$ is not observed for all units in the sample. This issue is known as sample selection, outcome attrition, or nonresponse. Let $S$ be a binary indicator variable such that $S_i=1$ if the outcome $Y_i$ is observed for unit $i$, and $S_i=0$ otherwise.  If the mechanism determining whether the outcome is observed ($S=1$) is related to the potential outcomes $Y(d)$ themselves, even after conditioning on treatment status $D$ and covariates $X$, then the subsample for whom we observe the outcome is no longer representative of the full population concerning the outcome process. Simply performing an analysis on the selected sample without accounting for the selection mechanism can introduce sample selection bias.
When non-random treatment assignment (violating $Y(d) \perp D$) occurs simultaneously with non-random sample selection (violating $Y(d) \perp S$), researchers face a so-called double selection problem, as emphasized by \citet{Bia2024}. In this situation, valid estimation requires assumptions addressing both sources of potential bias.
The first assumption invokes conditional independence of the treatment: 
\begin{assumption}{(Conditional Independence of the Treatment):}
\label{A1}
$$Y(d)\ \perp D \mid X = x \text{ for all } d \in \{0,1\} \text{ and } x \text{ in the support of } X.$$
\end{assumption}
By Assumption \ref{A1}, no unobservables jointly affect the treatment and the potential outcomes conditional on covariates $X$. 
Analogous to how Assumption \ref{A1} addresses confounding in treatment assignment, specific assumptions are required to handle sample selection. 

A common starting point is another selection-on-observables assumption, but applied to the selection process $S$. This is often termed the Missing-At-Random (MAR) assumption \citep{rubin1976inference} or conditional independence of selection. In the context of treatment evaluation, it means that conditional on observed variables (importantly, treatment status $D$ and covariates $X$), the selection indicator $S$ is independent of the potential outcomes:
\begin{align}\label{independenceselection}
Y(d) \perp S \mid D = d, X = x \quad \text{for all } d \in \{0,1\} \text{ and } x \text{ in the support of } X.
\end{align}
This conditional independence of
selection assumption states that, within groups defined by a specific treatment status $d$ and covariate values $x$, whether an outcome $Y(d)$ is observed ($S=1$) or missing ($S=0$) does not depend on the potential outcome's value itself. Selection is allowed to depend on treatment $D$ and covariates $X$, but not on any unobserved factors related to $Y(d)$ once $D$ and $X$ are accounted for.
However, this might be violated in many real-world scenarios. Selection could depend on unobserved factors (denoted $A$) that also influence the potential outcome, even after conditioning on $D$ and $X$. This is known as non-ignorable nonresponse or selection based on unobservables. In the following, we consider a much weaker condition, i.\,e., selection independence only holds if we could condition on the additional unobserved factors $A$. This leads to the following assumption:
\begin{assumption}{(Conditional Independence with Unobservables):}
\label{A4}
$$Y(d) \perp S \mid D = d, X = x, A = a \quad \text{for all } d \in \{0,1\} \text{ and } x,a \text{ in the support of } X \ \text{and}\ A.$$
\end{assumption}
Under Assumption \ref{A4}, selection is independent of potential outcomes once we account for treatment status, observed covariates, and the unobserved selection confounding factors $A$. While we cannot observe $A$ directly, we can apply the framework of \cite{chernozhukov2022long} to provide sharp bounds on the size of the omitted variable bias that results from not observing $A$. 
When introducing unobserved confounders $A$ in the selection process $S$, we need to specify how this affects the treatment assignment as well. We impose the following assumption:
\begin{assumption}{(No Unobserved Confounding in the Treatment Assignment):}\label{confoundinD}
$$A \perp D \mid X = x \quad \text{for all }  x \text{ in the support of } X. $$
\end{assumption}
It is worth noting that this assumption is implied by $X$ being the only causal parent of $D$, e.\,g., in a stratified randomized control trial. The analysis allowing unobserved confounders $A$ in both the sample selection and the treatment assignment will be pursued in an extended version of this work.

We also make the following assumption.
Let $p_d(X):=\mathbb{P}(D=d\mid X)$ for $d\in\{0,1\}$ and let
$\pi_0(d,X,A):=\mathbb{P}(S=1\mid D=d,X,A)$.

\begin{assumption}{(Common Support and Weak Overlap):}\label{A3} Assume (i) $p_d(X)>0$ and $\pi_0(d,X,A)>0$ almost surely for $d\in\{0,1\}$, and
(ii) the inverse-propensity moments satisfy
\begin{equation}\label{eq:weak-overlap}
\E\!\left[\frac{1}{p_1(X)\pi_0(1,X,A)}+\frac{1}{p_0(X)\pi_0(0,X,A)}\right] < \infty.
\end{equation}

\end{assumption}

We refer to Equation \eqref{eq:weak-overlap} as a \emph{weak overlap} condition since it requires only integrability of inverse propensities (rather than uniform lower bounds). The first part of the assumption is a conventional common support condition, which ensures that treatment assignment is non-degenerate and the probability of selection is always non-zero for each conditioning value. 

Further, we denote the conditional mean outcome by $\mu_d(X) = \mathbb{E}[Y|D=d, S=1, X] $. Under Assumption \ref{A1}, Assumption \ref{A3}, and conditional independence of selection in Equation \eqref{independenceselection}, the ATE is identified by:
$$
\theta_0 = \mathbb{E}[\phi_1 - \phi_0]
$$
with
\begin{align}\label{orthoscore}
\phi_d = \frac{\mathbf{1}\{D = d\} \cdot S \cdot [Y - \mu_d(X)]}{p_d(X) \cdot \pi_s(d, X)} + \mu_d(X)
\end{align}
being the efficient score function derived by \citet{Bia2024}. Hence, the ATE 
is identified using outcomes $Y$ from the selected sample ($S=1$) and selection indicators $S$ for all units.
Intuitively, identification involves modeling the conditional outcome mean within the selected sample, $\mathbb{E}[Y|D=d, S=1, X]$, and then appropriately adjusting or re-weighting based on estimates of the treatment propensity score $p_d(X) = \mathbb{P}(D=d|X)$ and the selection propensity score $\pi_s(d, X) = \mathbb{P}(S=1| D=d, X)$.

\section{Riesz Representers and Automatic Debiased Machine Learning} 

\subsection{Neyman-Orthogonal Scores and the Role of the Riesz Representer}

Many empirical problems now involve rich covariates. Machine-learning methods like Lasso, random forests, and neural networks can estimate nuisance functions such as conditional means and propensities in these settings. They achieve good prediction through regularization and model selection. These devices, however, typically introduce bias. If we plug a regularized estimate $\hat g$ into a target functional, the resulting estimator can inherit non-negligible bias and invalidate $\sqrt{n}$-consistent inference.

Debiased machine learning (DML) addresses this problem by using \emph{Neyman-orthogonal} scores \citep{Levit1975,IbragimovHasminskii1979,chernozhukov2018double}. In this framework, a score $\psi(W,\theta,g)$ identifies $\theta_0$ via the moment condition $$\E[\psi(W,\theta_0,g_0)]=0$$ 
and is constructed so that small errors in $g$ have only a second-order effect on the moment, where $W$ denotes the data. \citet{Bia2024} derive a Neyman-orthogonal score for high-dimensional sample selection models (see Equation \eqref{orthoscore}) and use cross-fitting to obtain valid inference.

A complementary approach uses the Riesz representer. For many parameters of interest (including the ATE), we can write
\[
\theta_0=\E[m(W,g_0)],
\]
where the map $g\mapsto \E[m(W,g)]$ is linear and continuous on a suitable function class. The Riesz Representation Theorem then yields a unique function $\alpha_0$, called the Riesz representer, such that
\[
\E[m(W,g)] = \E[\alpha_0(Z)\,g(Z)]
\]
for all admissible functions $g$, where $Z$ collects the arguments of $g$ and $\alpha_0$. 

The condition $\E[\alpha_0(Z)^2]<\infty$ is closely linked to $\theta_0$ having a finite semiparametric efficiency bound \citep{newey1994asymptotic, hirshberg2021augmented, chernozhukov2022automatic}. In our setting, the efficient score of \citet{Bia2024} admits an analogous Riesz representation that combines treatment and selection propensity weights.
The representer also leads to a generic orthogonal score. For target parameters of the form $\theta_0=\E[m(W,g_0)]$ with $g_0(Z)=\E[Y\mid Z]$, consider
\begin{align}\label{orthoscoreR}
\psi(W,\theta,g,\alpha)
= m(W,g)-\theta+\alpha(Z)\bigl(Y-g(Z)\bigr),
\end{align}
where $g$ and $\alpha$ approximate $g_0$ and $\alpha_0$. It is worth noting that in our sample selection model, $Y$ is only observed when $S=1$, so we basically consider $Y=SY$.
As shown by \citet{chernozhukov2022automatic}, evaluating at the true $\theta_0$ yields
\[
\E[\psi(W,\theta_0,g,\alpha)]
= -\,\E\bigl[(\alpha(Z)-\alpha_0(Z))(g(Z)-g_0(Z))\bigr].
\]
Thus, the score is doubly robust: its expectation is zero if either $g=g_0$ or $\alpha=\alpha_0$, and errors enter only through their product. Combined with cross-fitting, this property delivers $\sqrt{n}$-consistent inference with flexible first stages \citep{chernozhukov2018double}.

The Riesz formulation also plays an important role for estimation and sensitivity analysis. It casts the problem as learning a weighting function $\alpha_0$ jointly with $g_0$, which aligns well with variational, adversarial, and forest-based methods and can improve numerical stability and transparency. By learning the Riesz representer directly rather than relying on plug-in inverse probability weights, this approach can reduce instability when estimated propensities or selection probabilities are small. Most crucially for our case, the same representer-based structure naturally supports the sensitivity analysis in Section~\ref{sensitivity}.

\subsection{Riesz Representation Approach under Sample Selection}
Our goal is to identify the Average Treatment Effect (ATE), $\theta_0 = E[Y(1) - Y(0)]$, in the sample selection model described in Section \ref{sampleselectionmodel}, where
non-random treatment assignment occurs simultaneously with non-random
sample selection.  Under Assumptions \ref{A1}--\ref{A3}, the ATE admits the following representation in the
\emph{long model} (i.\,e., in a hypothetical setting where the latent factors $A$ were observed):
\begin{equation}\label{eq:theta0-long-rep}
\theta_0
= \E[m(W,g_0)]
= \E\!\left[g_0(1,X,A)-g_0(0,X,A)\right],
\end{equation}
where $W:=(Y,D,S,X,A)$ is the so-called long data vector and $g_0(d,x,a):=\E[Y\mid D=d,S=1,X=x,A=a]$ is the long regression.  Since $A$ is not observed in practice, we are only able to identify the so-called ``short" parameter
$$\theta_s = \mathbb{E}[m(W_s, g_s)]= \mathbb{E}[g_s(1, X) - g_s(0, X)]$$
from the observed short data vector $W_s:=(Y, D, S, X)$, where $g_s(d,X) = \mathbb{E}[Y| D=d, S=1, X]$ is the short regression. Since both parameters have a representation of the form $\theta=\mathbb{E}[m(W, g)]$, the Riesz Representation Theorem guarantees the existence of a Riesz representer $\alpha$, such that $\theta=\mathbb{E}[\alpha(Z) g(Z)]$.
The following main theorem of this paper, provides the explicit form of the Riesz representer in sample selection models.

 \begin{theorem}\label{Rieszrepresenter}
Under the Assumptions \ref{A1}, \ref{A4}, \ref{confoundinD} and \ref{A3}, the Riesz representers of the  long parameter $\theta_0$ and the short parameter $\theta_s$ are given by 
$$\alpha_0(w) = 
\frac{\mathbf{1}\{D=1\} \cdot S}{p_1(X) \pi_0(1, X, A)} 
- 
\frac{\mathbf{1}\{D=0\} \cdot S}{p_0(X) \pi_0(0, X, A)}$$
and
$$
\alpha_s(w) = 
\frac{\mathbf{1}\{D=1\} \cdot S}{p_1(X) \pi_s(1, X)} 
- 
\frac{\mathbf{1}\{D=0\} \cdot S}{p_0(X) \pi_s(0, X)},
$$
where $p_d(X) := \mathbb{P}(D = d| X)$ is the propensity score for $d \in \{0,1\}$, $\pi_0(d, X, A) = \mathbb{P}(S = 1| D = d, X, A)$ accounts for selection in the long parameter,
and $\pi_s(d, X) = \mathbb{P}(S = 1| D = d, X)$ accounts for selection in the short parameter.
 \end{theorem}
The formal proof is given in Appendix \ref{App:Rieszrepresenter}. Intuitively, the Riesz representer reweights the data to mirror what we would see in a randomized experiment. 
Weighting by $1/\mathbb{P}(D=d \mid X)$ increases the influence of units with observed characteristics $X$ that are unlikely to receive treatment $d$. This reweighting aligns the distribution of observed confounders, mimicking the balance achieved through random assignment.
Introducing sample selection creates an additional challenge: outcomes are observed only when $S=1$. To correct for this, we apply a second set of inverse-probability weights based on the likelihood of selection. Since we do not rely on conditional independence of selection in Equation \eqref{independenceselection} but rather on Assumption \ref{A4}, conditional independence of selection holds only after controlling for the unobserved variables $A$. 
As a result, our correction for selection must also account for these unobservables.
Weighting by $1/\pi_0(d,X,A)$ in the long parameter, or $1/\pi_s(d,X)$ in the short parameter, gives more weight to units that were less likely to be selected into the observed sample, thereby restoring representativeness relative to the full population.
The distinction between long and short parameters reflects whether the weighting scheme accounts for the unobserved confounders $A$ in the selection process $S$ or not.

 \subsection{Sensitivity Analysis} \label{sensitivity}
With observed data we are only able to identify the short parameter, although we are interested in the long parameter $\theta_0$. The Riesz representer theorem gives us a direct formula for the omitted variable bias arising from not controlling for $A$ in the selection into observability. Following \citet{chernozhukov2022long}, the difference between the long parameter $\theta_0$ and the short parameter $\theta_s$ is given by
$$\theta_0 - \theta_s = \mathbb{E}[(g_0 - g_s)(\alpha_0 - \alpha_s)],$$
which can be interpreted as the covariance between the error parts of $g$ and $\alpha$. 
Therefore, the (squared) bias is bounded by
$$|\theta_0 - \theta_s|^2 = \rho^2B^2 \leq B^2,$$
where $B^2:= \mathbb{E}[(g_0-g_s)^2]\mathbb{E}[(\alpha_0-\alpha_s)^2]$ and $\rho^2 := \text{Cor}^2(g_0-g_s,\alpha_0-\alpha_s)$. 
Furthermore, this squared bias bound \( B^2 \) has an intuitive decomposition that helps to understand the role of confounding in sample selection models. The squared bias bound \( B^2 \) can be decomposed as
\[
B^2 = \widetilde{S}^2 C^2_Y C^2_S,
\]
where $\widetilde{S}^2:= \mathbb{E}[(Y - g_s)^2] \mathbb{E}[\alpha_s^2]$, $C^2_Y := \frac{\mathbb{E}[(g_0 - g_s)^2]}{\mathbb{E}[(Y - g_s)^2]}$ and 
$C^2_S := \frac{\mathbb{E}[(\alpha_0 - \alpha_s)^2]}{\mathbb{E}[\alpha_s^2]}$. Therefore, the bound \( B^2 \) is the product of \( \widetilde{S}^2 \), a scaling factor identifiable from observed data, \( C^2_Y \) that measures confounding strength in the outcome equation and
\( C^2_S \) that measures confounding strength in the selection equation.
For \( C^2_Y \) and \( C^2_S \) researchers need to make informed assumptions about the impact of unobserved confounding.
More formally, it holds that
\begin{align*}
C^2_Y &= \frac{\mathbb{E}[(g_0-g_s)^2]}{\mathbb{E}[(Y-g_s)^2]} = R^2_{Y-g_s \sim g_0-g_s} = \eta^2_{Y \sim A \mid D,X,S=1},
\end{align*}
which measures the proportion of residual outcome variation (variation not explained by observed variables) that can be explained by the latent confounders $A$. It is by definition $\eta^2_{Y \sim A \mid D,X,S=1}$, the partial $R^2$ of $Y$ on the confounder $A$, after adjusting for $D$ and $X$, conditional on $S=1$.   
Further, it holds
\[
C_S^2=\frac{\mathbb{E}[\alpha_0^2]-\mathbb{E}[\alpha_s^2]}{\mathbb{E}[\alpha_s^2]}
=\frac{1-R^2_{\alpha_0\sim\alpha_s}}{R^2_{\alpha_0\sim\alpha_s}},
\quad \text{with} \quad
R^2_{\alpha_0\sim\alpha_s}=\frac{\mathbb{E}[\alpha_s^2]}{\mathbb{E}[\alpha_0^2]}.
\] 
It is worth noting that \( R^2_{\alpha_0 \sim \alpha_s} = \mathbb{E}[\alpha_s^2]/\mathbb{E}[\alpha_0^2] \) measures how much variation in the true Riesz representer \( \alpha_0 \) is explained by the short Riesz representer \( \alpha_s \). Therefore, \( 1 - R^2_{\alpha_0 \sim \alpha_s} \) (bounded between 0 and 1) measures the proportion of variation in \( \alpha_0 \) that is explained by the omitted confounder \( A \). 
While this parameter also admits an interpretation as a gain in precision, we find it more informative to use the following quasi-Gaussian approach for interpretation:
\\
\textbf{Quasi-Gaussian Selection Sensitivity.}
In practical applications, it might be difficult to think of plausible values for \( 1 - R^2_{\alpha_0 \sim \alpha_s}\), a technical and likely unfamiliar parameter. Instead, we find it useful to represent the selection indicator $S$ in a form of a latent index $S^*$ with Gaussian shocks crossing a threshold: Let $S=\mathbf{1}\{S^*>0\}$ with
$$
S^* = h(D,X) - U  \quad \text{and} \quad  U \mid D,X \sim N(0,1).$$
This representation does not entail loss of generality. We can then model confounding as follows:
$$
U = \mu_S A + \sqrt{1-\mu_S^2}\,\varepsilon_S, \quad \text{with} \quad  A,\varepsilon_S \stackrel{\mathrm{i.i.d.}}{\sim} N(0,1),
$$ 
independent of $(D,X)$.
Thus, $\mu^2_S$ is the $R^2$ in the regression of the Gaussian shock $U$ on the latent confounder $A$. By definition, it is also equal to $\eta^2_{S^* \sim A | D, X}$, the nonparametric partial $R^2$ in the regression of the latent index $S^*$ on $A$, after nonparametrically partialling out $(D,X)$. It is therefore easy to interpret. 
We can also map $\mu^2_S$ to the technical sensitivity parameter as follows. 
We compute the short selection probability
$\pi_s(d,x)=\mathbb{P}(S=1\mid D=d,X=x)=\Phi(h(d,x))$, so $h(d,x)=\Phi^{-1}(\pi_s(d,x))$ is identified from the short model,
and the long probability is
$$\pi_0(d,x,a)=\mathbb{P}(S=1\mid D=d,X=x,A=a)=\Phi\left((h(d,x)-\mu_S a)/\sqrt{1-\mu_S^2}\right).$$
We show in Appendix \ref{app:quasigaussian} that
$\mathbb{E}[\alpha_0^2]$ and $\mathbb{E}[\alpha_s^2]$ can be expressed in terms of these probabilities and can therefore be seen as functions of $\mu^2_S$. We then derive the maps from the interpretable to the technical sensitivity parameters:
$\mu^2_S \mapsto 
1-R^2_{\alpha_0\sim\alpha_s} (\mu^2_S).$
This yields a one-parameter, probit-scale
calibration of selection confounding that is directly compatible with the Riesz-based bias bounds.  Note that this does not impose any assumptions on the data, but is rather an interpretation device.
While sensitivity analysis maps assumptions about the unobserved confounder \(A\) (which might affect both outcome \(Y\) and selection \(S\)) to potential bias in the ATE estimate \(\theta_s\), it does not tell us how plausible those assumptions are. 
Researchers must therefore make informed judgments about the two partial $R^2$ measures that capture how strongly $A$  predicts the outcome $Y$ and the selection index $S^*$. 
This task can be aided by a benchmarking approach, following \citet{imbens2003sensitivity}, \citet{altonji2005selection}, \citet{oster2019unobservable}, \citet{cinelli2020making}, and \citet{chernozhukov2022long}, which uses the observed influence of specific covariates \(X_j\) as a reference point for the potential influence of an unobserved confounder $A$. We outline this approach in Appendix \ref{sec:benchmarking}.

\subsection{Estimation}\label{estimation}
Since \(\alpha_0\) is generally unknown, constructing a feasible estimator based on the orthogonal score (Equation \eqref{orthoscoreR}) in the DML framework requires an estimate \(\hat{\alpha}\). The traditional method for obtaining \(\hat{\alpha}\) is a plug-in approach. While conceptually straightforward, this plug-in approach for estimating the Riesz representer suffers from several drawbacks, particularly in high-dimensional or complex settings.
Deriving the analytical form of \(\alpha_0\) can be mathematically challenging or even intractable for more complex parameters of interest beyond the standard ATE. The formula for \(\alpha_0\) also frequently involves division by estimated probabilities or densities (see, e.\,g., $\hat{p}_d(X)$ and $\hat\pi_s(d,X)$ in Theorem \ref{Rieszrepresenter}). If these estimated quantities are close to zero, the resulting \(\hat{\alpha}\) can become extremely large. This occurs when the common support assumption (positivity) is empirically violated in the sample. Such large values can lead to unstable estimates of the target parameter \(\theta_0\). Recognizing the limitations of the plug-in method, recent research has focused on methods that
estimate the Riesz representer $\alpha_0$ directly, without needing its explicit analytical formula or relying on potentially unstable inverse weighting schemes. Two prominent direct approaches are variational methods (Riesz Regression) and adversarial (minimax) methods \citep{chernozhukov2020adversarial, chernozhukov2022automatic, chernozhukov2022riesznet}.
In this paper, we rely on the ForestRiesz, also developed by \cite{chernozhukov2022riesznet}, that adapts the random forest methodology to estimate the Riesz representer. Within this framework, the Riesz representer is modeled as locally linear with respect to a pre-specified feature map
$ a(Z) = \langle r(D, X, S), \beta(X) \rangle $,
where $Z=(D,X,S)$, $r(D, X, S)$ represents a smooth feature map (e.\,g., a polynomial series) and $\beta(X)$ denotes local coefficients that vary with covariates $X$. The algorithm constrains splits to covariates $X$ exclusively to preserve sufficient variation in the treatment variable $D$ within each node.
\cite{chernozhukov2022riesznet} show that this problem falls in the class of problems defined via solutions to moment equations $m(\cdot)=0$. Therefore, we can apply the framework of Generalized Random Forests of \cite{10.1214/18-AOS1709} to solve this local moment
problem via random forests.
For each node in the forest, the algorithm computes a Jacobian matrix and a local moment vector $$J(\text{node}) = \frac{1}{|\text{node}|}\sum_{i \in \text{node}} r(Z_i)r(Z_i)^\top \quad \text{and} \quad  M(\text{node}) = \frac{1}{|\text{node}|}\sum_{i \in \text{node}} m(W_i;r).$$ The optimal coefficient vector within each node is given by $\beta(\text{node}) = J(\text{node})^{-1} M(\text{node}).$
ForestRiesz grows the forest by recursively splitting nodes based solely on the covariates \(X\).
For each candidate split, the two resulting child nodes are evaluated by computing their respective
\(J\) and \(M\). The splitting rule seeks to maximize the stability-adjusted signal by minimizing
the aggregate local Riesz loss:
\[
  -\sum_{\text{child} \in \{1,2\}} |\text{child}| \; \beta(\text{child})^\top J(\text{child})\,\beta(\text{child}).
\]
This criterion favors splits that yield child nodes where the local moment \(M\) is both strong and
well-supported by a diverse (i.\,e., well-spread) feature set, while penalizing splits that produce
nodes with nearly singular \(J\).
ForestRiesz incorporates multitasking capabilities, wherein the forest simultaneously learns the regression function $\hat{g}$ and the Riesz representer $\hat\alpha$ by augmenting the node-splitting criteria with regression-based objectives. The final estimate is given by
\begin{align*}
    \hat{\theta}_{\mathrm{DR}}
= \mathbb{E}_n \big[\, m(W; \hat{g}) 
  + \hat{\alpha}(Z)\big(Y - \hat{g}(Z)\big) \,\big]
\end{align*}
or, better yet, its cross-fitted form to avoid overfitting,
leveraging Equation \eqref{orthoscoreR} as proposed in \cite{chernozhukov2022riesznet}, where $\mathbb{E}_n$ denotes the sample mean.

\section{Simulation Study}
The finite-sample properties of the proposed ForestRiesz (FR) estimator are assessed with a simulation study. The data-generative process (DGP) follows the conditional missing-at-random (MAR) design outlined in Appendix E of \cite{Bia2024}, with pre-treatment covariates $X$, a selection and treatment indicator $S, \thickspace D \in\{0,1\}$, 
error terms $u,\thickspace v,$ and $w$, and an outcome variable $Y$, that is only observed if $S = 1$:\\
$$
Y_i = \thickspace \theta_0 D_i \thickspace + X_i'\beta_0  + u_i, \quad
S_i = \thickspace \mathbf{1}\{D_i \thickspace  + \thickspace  X_i'\beta_0 \thickspace  + \thickspace  v_i \thickspace  > 0 \}, \quad
D_i = \mathbf{1}\{X_i'\beta_0 \thickspace  + \thickspace  w_i > 0\},
$$\\[-1.16 ex]
with $X_i \sim N(0,\sigma^2_X)$,  $(u_i,v_i) \sim N(0,\sigma^2_{u,v})$, and $w_i \sim N(0,1)$. 
For MAR to hold, $\sigma^2_{u,v}$ is specified as an identity matrix, implying that conditional on the treatment indicator and covariates none of the unobservables jointly affect the selection and outcome equation.  

In the DGP, we set the true ATE to $\theta_0 = 1$.
To benchmark the performance of the ForestRiesz, we compare it to an interactive regression model (IRM) \citep{chernozhukov2018double}, which does not adjust for the sample selection mechanism of the DGP, and to the sample selection model (SSM) by \cite{Bia2024}, which uses efficient Neyman-orthogonal score functions within the DML framework to address sample selection. 
The benchmark estimators are implemented via the \textit{doubleML} package \citep{Bach_DoubleML_-_Double}, using random forests\,\footnote{For the exact specification of hyperparameters of the random forests and the DML parameters see Appendix \ref{app_sub:sim_comp_details}.} 
for estimating the nuisance functions and three-fold cross-fitting to prevent overfitting bias.

For the sample sizes $N \in \{1000,\thickspace 4000,\thickspace 16000\}$,  Table \ref{tab:t_mc1} 
reports each estimator's average results across $200$ Monte Carlo iterations. For each estimator and sample size, it presents the estimate (ATE), the standard error (SE), and the corresponding bias (MAE). 
Across all sample sizes, the IRM model underestimates $\theta_0$, since it does not account for sample selection.
By contrast, both SSM and FR converge to the true $\theta_0=1$ when the number of observations increases. 
Moreover, as standard errors scale with $1/\sqrt{N}$, quadrupling the sample size reduces the standard errors of all estimators by approximately one half. 

\begin{table} [ht!]
\begin{center}
\begin{adjustbox}{max width=.97\textwidth}
\begin{threeparttable}
\begin{tabular}{c c ccc c ccc c ccc}
\hline \\[-2.5 ex]
&& \multicolumn{3}{c}{\textbf{IRM}} 
&& \multicolumn{3}{c}{\textbf{SSM}} 
&& \multicolumn{3}{c}{\textbf{FR}}
\\[.07 ex]
\hline\\[-2.5 ex]
N && ATE & SE & MAE
&& ATE & SE & MAE
&& ATE & SE & MAE\\[.07 ex]
\hline \\[-2.5 ex]
1000  
&& 0.8017 & 0.0564 & 0.1983
&& 1.1046 & 0.0451 & 0.1165 
&& 1.1306 & 0.0944 & 0.1365\\
4000 
&& 0.7457 & 0.0280 & 0.2543
&&1.0863 & 0.0222 & 0.0874
&& 1.0677 & 0.0461 & 0.0703\\
16000
&& 0.7046 & 0.0139 & 0.2954
&& 1.0621 & 0.0110 & 0.0622
&& 1.0349 & 0.0230 & 0.0357 \\[.16 ex]
\hline
\end{tabular}
\end{threeparttable}

%
%
%

\end{adjustbox}
\end{center}
\vspace{-.52 cm}
\caption{\scalebox{.952}{Average simulation results based on $\theta_0 = 1$ and $200$  Monte Carlo iterations.}}
\label{tab:t_mc1}
\end{table}

A more detailed comparison of the SSM and FR simulation results suggests a different bias-variance trade-off.
Across all sample sizes, SSM yields smaller standard errors, whereas FR results indicate a faster decline in bias as the sample size increases.
It is worth noting that the FR model is used without any tuning, while for the SSM we explored different random forest depths to improve propensity scores estimation and reduce bias.
To complement the previously described considerations, Figure \ref{fig:simulation} in Appendix \ref{app_sub:sim_add_res_ATE} presents the distribution of the ATE estimates 
across all Monte Carlo iterations.
Furthermore, Appendix \ref{app_sub:sim_add_res_SSM} presents additional results for the SSM estimator, showing that under the Lasso specifications used to learn the nuisance parameters in the score of \cite{Bia2024}, the SSM bias declines as expected given the linearity of the DGP. 

These considerations highlight the importance of the choice of machine learning methods and hyperparameter tuning in the SSM approach, and more generally within the DML framework (\cite{pmlr-v236-bach24a}), and demonstrate that the FR approach is considerably more robust.
A more detailed empirical comparison between the DML-based methods and the Riesz representer approach is left for future research.

\section{Application}

As an empirical application, we apply our method to estimate the gender wage gap in the U.\,S. We use data from the 2016 American Community Survey (ACS), which provides a representative 1\,\% sample of the U.\,S. population under mandatory participation. Since some respondents do not report their wages, even though they are employed, any gender wage gap analysis based on the ACS data is subject to a sample selection problem. The dataset contains 158 variables for socio-economic characteristics at the individual and the household level, for example referring to education, industry, and occupation. 
We follow the study of \cite{bach2024heterogeneity} and focus on two sub-populations in the ACS: respondents with a high school degree and those with a college degree. Our treatment variable $D$ is the gender of a respondent, with $D=1$ indicating a female respondent. Our outcome variable $Y$ denotes (log) weekly wages (in USD) and the indicator $S$ indicates whether $Y$ is observed (i.\,e., the respondent has reported her wage). In the high school sub-population, we have $372\ 728$ respondents and in the college sub-population $297\ 178$ individuals.

In order to estimate the gender wage gap, we apply the proposed ForestRiesz, where one fits a random forest that jointly learns the Riesz representer $\alpha$ and the regression function $g$ in one step as described in Section \ref{estimation}. 
To demonstrate the relevance of our Riesz representer approach in sample selection models, we compare our estimation results with those obtained from the interactive regression model (IRM)  and the SSM approach, both implemented using the \textit{doubleML} package \citep{Bach_DoubleML_-_Double}, as in the simulation study. We apply the three estimators to the high school and college subsamples and report point estimates, standard errors, and p-values.
Table \ref{tab:est_res} presents the estimation results for the college and the high school subsamples. For all three regression models, we find a significant gender wage gap in both subsamples, with a larger gap in the high school subsample than in the college subsample, in line with previous findings in \cite{bach2024heterogeneity}. Since the estimated wage gap is approximately $3$ percentage points larger using the Riesz representer approach compared to IRM, our results suggest that we underestimate the gender wage gap when not controlling for non-reporting respondents. Applying a logit model to the reporting indicator $S$, we find that never-married female workers with a high university degree (professional degree) have a higher probability of reporting their income than their male counterparts, and that the relationship between experience and reporting also differs between men and women (see Table \ref{tab:reporting} in Appendix \ref{sec:tab_fig}). Because these covariates are also among the strongest predictors of wages (see Table \ref{tab:predictors} in Appendix \ref{sec:tab_fig}), estimates of the gender wage gap are subject to selection bias if these patterns are ignored. While the IRM model does not address this issue, both the ForestRiesz (FR) and the SSM approach correct for it by reweighting respondents with a lower probability of wage reporting.

\begin{table}[ht!]
\begin{center}
\begin{adjustbox}{max width=.97\textwidth}

\begin{threeparttable}
\begin{tabular}{l c cc c cc c cc}
\hline \\[-2.5 ex]
&& \multicolumn{2}{c}{\textbf{IRM}} 
&& \multicolumn{2}{c}{\textbf{SSM}} 
&& \multicolumn{2}{c}{\textbf{FR}}
\\[.07 ex]
\hline\\[-2.5 ex]
&& College & High school 
&& College & High school 
&& College & High school \\[.07 ex]
\hline \\[-2.5 ex]
Estimate  
&& -0.0989{***} & -0.141{***} 
&& -0.153{***} & -0.198{***} 
&&  -0.128{***} & -0.172{***} \\
SE 
&& 0.003 & 0.003 
&& 0.001 & 0.001  
&& 0.002 & 0.002  \\
P-value
&& 0.000   & 0.000  
&& 0.000   & 0.000  
&& 0.000   & 0.000\\[.16 ex]
\hline
\end{tabular}
\end{threeparttable}
 
\end{adjustbox}
\end{center}
\vspace{-.43 cm}
\caption{\scalebox{.952}{Estimation results for gender wage gap. Significance: \scalebox{.97}{*** $p<0.01$, ** $p<0.05$, * $p<0.10$.}}}
\label{tab:est_res}
\end{table}

Next, we conduct a sensitivity analysis to assess the robustness of our estimated treatment effects to unobserved confounding. Using observed covariates $X_j$ as benchmarks, this approach evaluates how influential an unobserved confounder $A$ would need to be to overturn our main findings. We perform this analysis for all covariates and report results for the six most influential covariate groups in the college subsample in Table \ref{tab:bench} in Appendix \ref{sec:tab_fig}.
For each group $j$, the table reports
the share of additional outcome variation $G_{Y,j}$, selection variation $G_{S,j}$, and their alignment measure $\rho_j$, detailed in Appendix \ref{sec:benchmarking}.
Overall, the results indicate that the estimated gender wage gap in the college subsample is highly robust. Omitting the most influential covariate group, marital status, changes the ATE estimate by only $0.55$ percentage points. 
Notably, although education explains the largest share of variation in wages and in the Riesz representer (high $G_Y$ and $G_S$), it has virtually no effect on the estimated gender wage gap (low $\Delta \theta$), reflecting the weak correlation between the residual component of the outcome and Riesz representer models (small $|\rho|$).
We further assess robustness through sensitivity analyses based on these benchmarks.
First, we construct confidence intervals that account for unobserved confounding as strong as the marital status covariate.
Figure \ref{fig:CI} in Appendix \ref{sec:tab_fig} shows that even under this conservative scenario, the estimated ATE remains statistically significant.
Second, we examine the magnitude of unobserved confounding required to overturn our conclusions.
Figure \ref{fig:sensitivity} in Appendix \ref{sec:tab_fig} illustrates the potential bias as a function of $C^2_Y = \eta^2_{Y \sim A|D,X,S=1}$ and $\eta^2_{S^*\sim~ A |D, X}$, assuming the worst-case alignment ($\rho=1$).
The robustness value (RV) for the college subsample is $0.063$, implying that an unobserved confounder would need to explain at least $6.3\,\%$ of both residual outcome and selection variation to nullify the estimated effect. This is substantially more than any observed covariate in our data can explain.

\section{Conclusion}One main contribution of the paper is a bounds analysis for treatment effects when the traditional sample-selection model’s conditional missing-at-random (MAR) assumption fails. Although MAR is widely used, it is often hard to defend in applications. We relax MAR by introducing a latent confounder that affects selection and then derive the Riesz representer for the average treatment effect (ATE), which combines treatment-propensity weighting with selection-probability weighting. Using the resulting Riesz representers for the short and long models, we decompose the omitted-variable bias into three interpretable components. This decomposition yields sharp, distribution-free bounds on the magnitude of bias and provides a practical sensitivity-analysis toolkit for violations of MAR.

A second contribution is to adapt the ForestRiesz method of \citet{chernozhukov2022riesznet} to treatment-effect estimation under sample selection. This automatic debiased machine learning approach jointly learns the outcome regression and the Riesz representer, avoiding the numerical instability of plug-in estimators that require direct inversion of estimated probabilities. Our simulations highlight the advantages of the ForestRiesz framework over more standard doubly robust plug-in approaches in finite samples. We illustrate the practical benefits of the method in an application to the U.\,S. gender wage gap using the American Community Survey. We find that ignoring sample selection leads to an underestimation of the wage gap, driven by systematic gender differences in wage reporting. A benchmarking-based sensitivity analysis indicates that this conclusion is robust.

Overall, our results highlight the importance of explicitly accounting for sample selection, particularly in survey-based studies, and demonstrate that the ForestRiesz estimator offers a robust, interpretable, and computationally attractive approach for causal inference in the presence of selective outcome observability.

\clearpage

\bibliography{references}

\clearpage

\appendix

\section{Proof of Theorem \ref{Rieszrepresenter}}\label{App:Rieszrepresenter}

We derive the result for the long parameter, as the proof for the short parameter is analogous.
We aim to show that
\[
\theta_0=\E[Y(1)-Y(0)] = \E[m(W,g_0)] = \E[g_0(D,X,A)\alpha_0(W)]
\]
with $g_0(d,x,a):=\E[Y\mid D=d,S=1,X=x,A=a]$ and $
m(W,g_0):=g_0(1,X,A)-g_0(0,X,A)$.
\\ \\
\noindent\textbf{Step 1:} First, we show that $\theta_0=\E[m(W,g_0)]$. It suffices to show that, for each $d\in\{0,1\}$, it holds that
\[
\E[g_0(d,X,A)] = \E[Y(d)].
\]
Fix $d\in\{0,1\}$. Then,
\begin{align*}
\E[g_0(d,X,A)]
&= \E\!\left[\E\!\left[Y \mid D=d,S=1,X,A\right]\right] \\
&= \E\!\left[\E\!\left[Y(d) \mid D=d,S=1,X,A\right]\right] 
&& \text{(Observational Rule)}\\
&= \E\!\left[\E\!\left[Y(d) \mid D=d,X,A\right]\right]
&& \text{(Assumption \ref{A4})}\\
&= \E\!\left[\E\!\left[\E\!\left[Y(d) \mid D=d,X,A\right]\mid X\right]\right]
&& \text{(Law of Iterated Expectation)}\\
&= \E\!\left[\E\!\left[\E\!\left[Y(d) \mid D=d,X,A\right]\mid D=d,X\right]\right]
&& \text{(Assumption \ref{confoundinD})}\\
&= \E\!\left[\E\!\left[Y(d)\mid D=d,X\right]\right]
&& \text{(Law of Iterated Expectation)}\\
&= \E\!\left[\E\!\left[Y(d)\mid X\right]\right]
&& \text{(Assumption \ref{A1})}\\
&= \E[Y(d)].
\end{align*}
Therefore,
\[
\E[m(W,g_0)]
= \E[g_0(1,X,A)]-\E[g_0(0,X,A)]
= \E[Y(1)]-\E[Y(0)]
= \theta_0.
\]

\medskip
\noindent\textbf{Step 2:} \emph{Verify the Riesz representer.}
Define
\[
\alpha_0(W)
:= \frac{\mathbf{1}\{D=1\}S}{p_1(X)\pi_0(1,X,A)}
 - \frac{\mathbf{1}\{D=0\}S}{p_0(X)\pi_0(0,X,A)}.
\]
It holds that
\begin{align*}
\E[\alpha_0(W)\,g_0(D,X,A)]
&= \E\!\left[\E[\alpha_0(W)\,g_0(D,X,A)\mid X,A]\right]\\
&= \E\!\left[
g_0(1,X,A)\frac{\E[\mathbf{1}\{D=1\}S\mid X,A]}{p_1(X)\pi_0(1,X,A)}
-
g_0(0,X,A)\frac{\E[\mathbf{1}\{D=0\}S\mid X,A]}{p_0(X)\pi_0(0,X,A)}
\right]\\
&= \E\!\left[g_0(1,X,A)-g_0(0,X,A)\right]
= \E[m(W,g_0)],
\end{align*}
where we used that
\begin{align*}
\E[\mathbf{1}\{D=d\}S\mid X,A]
&= \mathbb{P}(D=d,S=1\mid X,A)\\
&= \mathbb{P}(D=d\mid X,A)\,\mathbb{P}(S=1\mid D=d,X,A)
= p_d(X)\,\pi_0(d,X,A),
\end{align*}
with $\mathbb{P}(D=d\mid X,A)=p_d(X)$ by Assumption \ref{confoundinD}.

\section{Omitted Variable Bias in Sample Selection Models}\label{proofsensitivity}
Here, we apply the framework of \cite{chernozhukov2022long} to derive the omitted variable bias in the sample selection model with confounding in selection.
Let $\theta_0$ denote the long parameter and $\theta_s$ the short parameter,
\[
\theta_0 = \E[m(W,g_0)] \quad and \quad \theta_s = \E[m(W_s,g_s)],
\]
where $g_0$ and $g_s$ are the long and short outcome regressions defined in the main text.
Let $\alpha_0$ and $\alpha_s$ be the corresponding long and short Riesz representers. The omitted variable bias (OVB) admits the representation
\begin{align}\label{eq:ovb-inner-product}
\theta_0-\theta_s \;=\; \E\big[(g_0-g_s)(\alpha_0-\alpha_s)\big].
\end{align}

As described in the main text, it holds that
\begin{align}
|\theta_0-\theta_s|^2 = \rho^2 B^2 \le B^2   
\end{align}
with 
\begin{equation}\label{eq:B-decomp}
B^2 = \widetilde S^{\,2}\, C_Y^2\, C_S^2,
\end{equation}
where $\widetilde S^{\,2}$ is identified from the observed data, while $C_Y^2$ and
$C_S^2$ summarize the strength of omitted-variable effects in the outcome and selection
components, respectively. In particular,

\begin{align*}
C^2_Y &= \frac{\mathbb{E}[(g_0-g_s)^2]}{\mathbb{E}[(Y-g_s)^2]} = R^2_{Y-g_s \sim g_0-g_s} = \eta^2_{Y \sim A \mid D,X,S=1}
\end{align*}
is the fraction of residual outcome variation (after controlling for observed covariates) that
is explained by the omitted confounder through the long regression.

Next, we consider the sensitivity parameter $C_S^2$ in more detail.
Let $\mathcal{A}_s$ be the closed linear subspace of $L^2(\mathbb{P})$ consisting of square-integrable
functions measurable with respect to the short information set (the observed variables in $W_s$).
Since the long functional restricted to $\mathcal{A}_s$ has the Riesz representer $\alpha_s$,
we have
\[
\E[(\alpha_0-\alpha_s)a]=0 \quad \text{for all } a\in\mathcal{A}_s,
\]
so $\alpha_s$ is the $L^2$-projection of $\alpha_0$ onto $\mathcal{A}_s$. Taking $a=\alpha_s$
yields $\E[\alpha_0\alpha_s]=\E[\alpha_s^2]$, and hence 
\begin{equation}\label{eq:pythag}
\E[(\alpha_0-\alpha_s)^2] = \E[\alpha_0^2]-\E[\alpha_s^2].
\end{equation}
Therefore,
\begin{equation}\label{eq:CsR2}
C_S^2
= \frac{\E[\alpha_0^2]-\E[\alpha_s^2]}{\E[\alpha_s^2]}
= \frac{1-R^2_{\alpha_0\sim\alpha_s}}{R^2_{\alpha_0\sim\alpha_s}},
\qquad  \text{with}\qquad
R^2_{\alpha_0\sim\alpha_s}:=\frac{\E[\alpha_s^2]}{\E[\alpha_0^2]}.
\end{equation}
The quantity $1-R^2_{\alpha_0\sim\alpha_s}$ measures the share of variation in the long
representer that is not captured by the short representer. Next, we consider the closed-form expressions for the Riesz representers.

\subsection*{Closed-form expressions for $\E[\alpha_0^2]$ and $\E[\alpha_s^2]$}
The long Riesz representer is given by
\[
\alpha_0(W)
:= \frac{\1\{D=1\}S}{p_1(X)\pi_0(1,X,A)}
 - \frac{\1\{D=0\}S}{p_0(X)\pi_0(0,X,A)},
\]
and the short Riesz representer by
\[
\alpha_s(W_s)
:= \frac{\1\{D=1\}S}{p_1(X)\pi_s(1,X)}
 - \frac{\1\{D=0\}S}{p_0(X)\pi_s(0,X)}.
\]
Because $\1\{D=1\}\1\{D=0\}=0$, the cross term vanishes and therefore
\[
\E[\alpha_0^2] = \E\!\left[\Big(\tfrac{\1\{D=1\}S}{p_1(X)\pi_0(1,X,A)}\Big)^2\right]
              + \E\!\left[\Big(\tfrac{\1\{D=0\}S}{p_0(X)\pi_0(0,X,A)}\Big)^2\right].
\]
For $d\in\{0,1\}$, we have
\begin{align*}
\E\!\left[\Big(\tfrac{\1\{D=d\}S}{p_d(X)\pi_0(d,X,A)}\Big)^2 \,\Big|\, X,A\right]
&= \frac{\E[\1\{D=d\}S\mid X,A]}{p_d(X)^2\pi_0(d,X,A)^2}\\
&= \frac{p_d(X)\pi_0(d,X,A)}{p_d(X)^2\pi_0(d,X,A)^2}
= \frac{1}{p_d(X)\pi_0(d,X,A)}.
\end{align*}
Hence,
\begin{equation}\label{eq:alpha0-second-moment}
\E[\alpha_0^2]
= \E\!\left[\frac{1}{p_1(X)\pi_0(1,X,A)}+\frac{1}{p_0(X)\pi_0(0,X,A)}\right].
\end{equation}
Similarly, we can show that
\begin{equation}\label{eq:alphas-second-moment}
\E[\alpha_s^2]
= \E\!\left[\frac{1}{p_1(X)\pi_s(1,X)}+\frac{1}{p_0(X)\pi_s(0,X)}\right].
\end{equation}
Combining the Equations \eqref{eq:pythag}--\eqref{eq:alphas-second-moment} gives
\begin{equation}\label{eq:Cs-closed-form}
C_S^2
=
\frac{
\E\!\left[\frac{1}{p_1(X)\pi_0(1,X,A)}+\frac{1}{p_0(X)\pi_0(0,X,A)}\right]
-
\E\!\left[\frac{1}{p_1(X)\pi_s(1,X)}+\frac{1}{p_0(X)\pi_s(0,X)}\right]
}{
\E\!\left[\frac{1}{p_1(X)\pi_s(1,X)}+\frac{1}{p_0(X)\pi_s(0,X)}\right]
}.
\end{equation}
The terms $1/(p_d(X)\pi(\cdot))$ grow when either the treatment propensity $p_d(X)$ or the
selection probability $\pi(\cdot)$ is small. Thus, $\E[\alpha_0^2]$ and $\E[\alpha_s^2]$ summarize the overlap and
selection difficulty through an average inverse-probability scale. Consequently, the sensitivity parameter $C_S^2$ in Equation
\eqref{eq:Cs-closed-form} measures how much the representer varies when the
selection model does or does not depend on the unobserved confounder $A$, and it can be interpreted as the gain in precision from observing $A$.
The Riesz Representer Framework requires that $\E[\alpha_0^2]<\infty$ and $\E[\alpha_s^2]<\infty$.
A convenient sufficient condition is
\[
\E\!\left[\frac{1}{p_1(X)\pi_0(1,X,A)}+\frac{1}{p_0(X)\pi_0(0,X,A)}\right]<\infty,\
\E\!\left[\frac{1}{p_1(X)\pi_s(1,X)}+\frac{1}{p_0(X)\pi_s(0,X)}\right]<\infty,
\]
which we refer to as a \emph{weak overlap} condition as stated in Assumption \ref{A3}.

\section{Quasi-Gaussian Latent-Index Model for Selection}\label{app:quasigaussian}

This section provides an interpretable calibration of $C_S^2$ using a probit-style latent-index
model for selection. The model serves purely as a calibration device and is not required for the identification results presented in the main text.

\subsection{Latent-Index Specification}
Assume the long selection mechanism admits the representation
\[
S=\1\{S^*>0\},\qquad
S^* = h(D,X) - U,\qquad
U = \mu_S A + \sqrt{1-\mu_S^2}\,\varepsilon_S,
\]
where $A,\varepsilon_S\stackrel{i.i.d.}{\sim}N(0,1)$ and $(A,\varepsilon_S)\perp (D,X)$.
Then,
\[
\pi_s(d,x)=\P(S=1\mid D=d,X=x)=\Phi(h(d,x)),
\qquad
h(d,x)=\Phi^{-1}(\pi_s(d,x)),
\]
and
\[
\pi_0(d,x,a)=\P(S=1\mid D=d,X=x,A=a)
=\Phi\!\left(\frac{h(d,x)-\mu_S a}{\sqrt{1-\mu_S^2}}\right).
\]
\noindent The scalar $\mu_S^2\in[0,1)$ is the latent partial $R^2$ of $A$ in the selection index, that is:
\[
\mu_S^2 = R^2_{S^*\sim A\mid D,X}.
\]
Under the normalization $\Var(U\mid D,X)=1$, we have $\Var(S^*\mid D,X)=1$ and
$$\Var(\E[S^*\mid D,X,A]\mid D,X)=\Var(\mu_S A)=\mu_S^2.$$

\subsection{Mapping $\mu_S^2$ to $C_S^2$ and $R^2_{\alpha_0\sim\alpha_s}$}
Given $(p_d,\pi_s)$ and a choice of $\mu_S^2\in[0,1)$, define $h(d,x)=\Phi^{-1}(\pi_s(d,x))$ and
\[
\pi_0(d,x,a;\mu_S^2)
:=\Phi\!\left(\frac{h(d,x)-\sqrt{\mu_S^2}\,a}{\sqrt{1-\mu_S^2}}\right).
\]
Then, we can express the Riesz representer $\alpha_0$ as a function of $\mu_S^2$:
\[
\E[\alpha_0^2(\mu_S^2)]
=
\E\!\left[\frac{1}{p_1(X)\pi_0(1,X,A;\mu_S^2)}+\frac{1}{p_0(X)\pi_0(0,X,A;\mu_S^2)}\right].
\]
The resulting calibration curve is given by
\[
C_S^2(\mu_S^2)
=\frac{\E[\alpha_0^2(\mu_S^2)]-\E[\alpha_s^2]}{\E[\alpha_s^2]}
=\frac{\E[\alpha_0^2(\mu_S^2)]}{\E[\alpha_s^2]}-1.
\]
Equivalently,
\[
R^2_{\alpha_0\sim\alpha_s}(\mu_S^2)=\frac{\E[\alpha_s^2]}{\E[\alpha_0^2(\mu_S^2)]},
\quad \text{and}\quad
C_S^2(\mu_S^2)=\frac{1-R^2_{\alpha_0\sim\alpha_s}(\mu_S^2)}{R^2_{\alpha_0\sim\alpha_s}(\mu_S^2)}.
\]

\subsection{Practical Computation}

Let $\widehat p_d(X_i)$ and $\widehat\pi_s(d,X_i)$ be estimates from the observed data.
For a grid of $\mu_S^2$ values, we perform the following steps:
\begin{enumerate}
\item Compute $\widehat h(d,X_i)=\Phi^{-1}(\widehat\pi_s(d,X_i))$.
\item Draw $A_i^{(b)}\sim N(0,1)$ independently for $b=1,\dots,B$, and compute
\[
\widehat\pi_0^{(b)}(d,X_i)
=\Phi\left(\frac{\widehat h(d,X_i)-\sqrt{\mu_S^2}\,A_i^{(b)}}{\sqrt{1-\mu_S^2}}\right).
\]
\item Approximate $\E[\alpha_0^2(\mu_S^2)]$ and $\E[\alpha_s^2]$ by
\[
\widehat{\E}[\alpha_0^2(\mu_S^2)]
=
\frac{1}{nB}\sum_{i=1}^n\sum_{b=1}^B
\left(\frac{1}{\widehat p_1(X_i)\widehat\pi_0^{(b)}(1,X_i)}
+\frac{1}{\widehat p_0(X_i)\widehat\pi_0^{(b)}(0,X_i)}\right)
\]
and
\[
\widehat{\E}[\alpha_s^2]
=
\frac{1}{n}\sum_{i=1}^n
\left(\frac{1}{\widehat p_1(X_i)\widehat\pi_s(1,X_i)}
+\frac{1}{\widehat p_0(X_i)\widehat\pi_s(0,X_i)}\right).
\]
\item Report $\widehat C_S^2(\mu_S^2)=\widehat{\E}[\alpha_0^2(\mu_S^2)]/\widehat{\E}[\alpha_s^2]-1$,
or $\widehat R^2_{\alpha_0\sim\alpha_s}(\mu_S^2)=\widehat{\E}[\alpha_s^2]/\widehat{\E}[\alpha_0^2(\mu_S^2)]$.
\end{enumerate}
\ \\
The following figure provides a graphical illustration of the computation:
\begin{figure}[ht!]
  \centering
\includegraphics[width=0.65\linewidth]{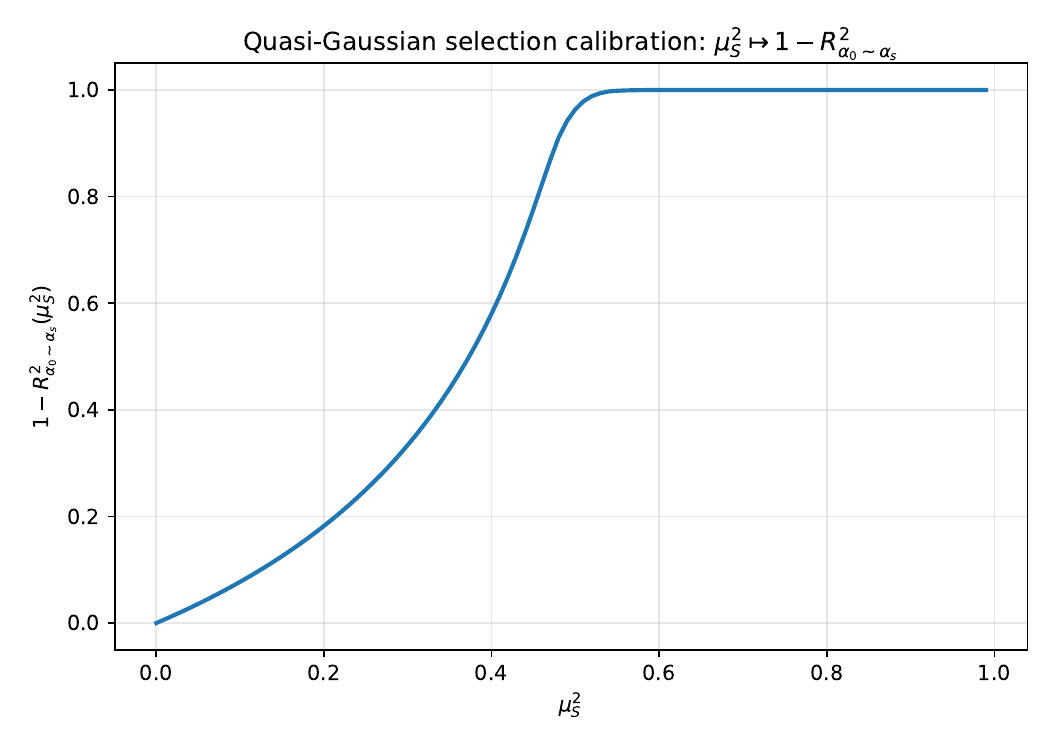}\\[-.35cm]
  \caption{Quasi-Gaussian calibration curve in a synthetic example. The horizontal axis shows values of the classical interpretable parameter, and the vertical axis shows values of the implied technical parameter.}
  \label{fig:qg-calib}
\end{figure}

\newpage
\section{Benchmarking Sensitivity to Unobserved Confounding}
\label{sec:benchmarking}

Relying on benchmarking, we measure how much a specific observed variable \(X_j\) actually matters in our data by looking at its influence in four key areas. Let \(g_s\) and \(\alpha_s\) be the outcome model and the Riesz representer using all covariates \(X\), and let \(g_{s,-j}\) and \(\alpha_{s,-j}\) be the versions omitting \(X_j\). We consider four quantities to measure the impact of the omitted variable $X_j$:

\begin{enumerate}
    \item \textbf{Outcome Prediction:} We measure \(X_j\)'s impact on predicting the outcome \(Y\) (within the selected sample, \(S=1\)) by calculating the increase in R-squared (\(\Delta \eta^2_{Y \sim X_j | D,X_{-j},S=1} := \eta^2_{Y \sim D,X,S=1} - \eta^2_{Y \sim D,X_{-j},S=1}\)) when \(X_j\) is added to the model. This shows how much \(X_j\) improves outcome prediction beyond other variables.

    \item \textbf{Selection Weights:} We measure \(X_j\)'s impact on the statistical weights \(\alpha_s\) used for correction by calculating the relative change in the weights' overall size (\(1 - R^2_{\alpha_s \sim \alpha_{s,-j}} := (\mathbb{E}[\alpha_s^2] - \mathbb{E}[\alpha_{s,-j}^2]) / \mathbb{E}[\alpha_s^2]\)) when \(X_j\) is included. This shows how much \(X_j\) changes the necessary adjustment for selection and treatment assignment.

    \item \textbf{ATE Estimate:} We measure \(X_j\)'s direct impact on the final result by calculating the change in the ATE estimate (\(\Delta \theta_{s,j} := \theta_{s,-j} - \theta_s\)) when \(X_j\) is included versus excluded as a control variable. This shows how sensitive the estimated ATE is to controlling for \(X_j\).

    \item \textbf{Alignment of Effects:} We measure whether \(X_j\)'s effects on the outcome and selection weights work together or against each other by calculating the correlation (\(\rho_j := \text{Cor}(g_{s,-j} - g_s, \alpha_s - \alpha_{s,-j})\)) between the changes they cause when \(X_j\) is removed.
    
\end{enumerate}

Then, we calculate the following three metrics for \(X_j\) to define benchmark values for the sensitivity parameters:

\begin{enumerate}
    \item \textbf{Outcome Gain Metric (\(G_{Y,j}\)):} This serves as a benchmark for how much \(A\) might explain the remaining variance in the outcome \(Y\) (after accounting for \(D, S=1,\) and \(X\)). Hence, it is a proxy for the sensitivity parameter \(C^2_Y = \eta^2_{Y \sim A | D,X,S=1}\), the partial $R^2$ of $Y$ on the confounder $A$. The assumption is that \(A\)'s relative contribution to explaining residual outcome variance is similar to \(X_j\)'s:
    \[
    G_{Y,j} := \frac{\Delta \eta^2_{Y \sim X_j | D,X_{-j},S=1}}{1 - \eta^2_{Y \sim D,X,S=1}} \approx C^2_Y = \eta^2_{Y \sim A|D,X,S=1}.
    \]
    \textit{Interpretation:} If \(X_j\) explains, say, 5\,\% of the outcome variance that was previously unexplained by \(D, S=1,\) and \(X_{-j}\) (resulting in \(G_{Y,j} = 0.05\)), this sets a benchmark. We can then ask: "Is it plausible that the unobserved confounder \(A\) explains more than 5\,\% of the residual outcome variance?" This directly informs the choice of \(C^2_Y\) in the sensitivity analysis.

    \item \textbf{Selection / Representer Gain Metric (\(G_{S,j}\)):} This serves as a benchmark for \(A\)'s association with the selection mechanism, captured by the sensitivity parameter \(C^2_S = (1 - R^2_{\alpha_0 \sim \alpha_s}) / R^2_{\alpha_0 \sim \alpha_s}\) or $1 - R^2_{\alpha_0 \sim \alpha_s}$, respectively. Therefore, we link the relative change in the Riesz representer due to \(A\) to the change in the Riesz representer due to the observed \(X_j\):
    \[
    G_{S,j} := 1 - R^2_{\alpha_s \sim \alpha_{s,-j}} \approx 1 - R^2_{\alpha_0 \sim \alpha_s}.
    \]
    \textit{Interpretation:} $G_{S,j}$ quantifies how strongly $X_j$ influences the selection mechanism (encoded in $\alpha_s$), setting a benchmark for the magnitude of $A$'s impact. Higher $G_{S,j}$ values imply a higher threshold for $A$'s assumed effect. 

    \item \textbf{Correlation / Degree of Adversity Metric (\(\rho_j\)):} This metric captures how aligned the confounding effects of \(X_j\) are on the outcome and selection mechanism (via the RR). It measures the correlation between the change in the outcome model \(g_s\) and the change in the Riesz representer \(\alpha_s\) when \(X_j\) is removed:
    \[
    \rho_j := \text{Cor}(g_{s,-j} - g_s, \alpha_s - \alpha_{s,-j}).
    \]
    \textit{Interpretation:} \(\rho_j\) reflects alignment of \(X_j\)'s confounding effect. A value close to +1 or -1 indicates that \(X_j\) influences both the outcome prediction (within the selected sample) and the selection mechanism representation \(\alpha_s\) in a similar way, leading to a larger change in the ATE estimate (larger \(\Delta \theta_{s,j}\)). We can compare the assumed \(\rho\) for \(A\) against the observed \(\rho_j\) for plausible observed confounders \(X_j\).
\end{enumerate}
Calculating \(G_{Y,j}\), \(G_{S,j}\), and \(\rho_j\) for one or more carefully chosen covariates \(X_j\) provides concrete reference points. These points correspond directly to values used in the sensitivity analysis (\(C_Y^2\), \(C_S^2\) and \(\rho\)). They help evaluate whether overturning the study's main conclusions would require the unobserved confounder \(A\) to be substantially more influential (in terms of outcome variance explained, impact on the selection mechanism's RR structure, or correlation/adversity) than key observed covariates like \(X_j\).

\clearpage

\section{Additional Material for the Simulation Study}
\label{app:sim}

\subsection{Computational Details}

\label{app_sub:sim_comp_details}
\begin{table} [ht!]
\begin{center}
\begin{adjustbox}{max width=.79\textwidth}
\begin{threeparttable}
	\begin{tabular}{lll}
        \hline \\[-1.06ex]
        \textbf{Parameters - scikit-learn} & & 
        \textbf{Parameters - doubleML}\\ [.07ex]
        \hline 
        \hline\\[-1.6 ex]
        \underline{RandomForest classes}   && \underline{IRM and SSM} \\
        &&  \hspace{.25 cm} n\_folds$=3$, \thickspace n\_rep$=1$ \\[-1.96ex]
        \hspace{.25 cm} n\_estimators $=500$ &&  \\
        \hspace{.25 cm} max\_depth $=20$  && \underline{SSM}\\
        \hspace{.25 cm} min\_samples\_leaf $=5$ && \hspace{.25 cm} score = 'missing-at-random'\\
        \hspace{.25 cm} max\_features $=$"sqrt"&& \hspace{.25cm} normalize\_ipw = True\\[.43ex]
        \hline 
    \end{tabular}
\end{threeparttable} 
\end{adjustbox}
\end{center}
\vspace{-.43 cm}
\caption{This table reports 
the final hyperparameter set up used for the \textit{RandomForestRegressor} and \textit{RandomForestClassifier} classes from \textit{scikit-learn} \citep{pedregosa2011scikit}, as well as the settings for the estimator classes \textit{DoubleMLIRM} and \textit{DoubleMLSSM} from the \textit{doubleML} \citep{Bach_DoubleML_-_Double} Python package. Parameters not reported are kept at their default values.}
\label{tab:comp_details}
\end{table}

\subsection{Additional Simulation Results: ATE Estimates}
\label{app_sub:sim_add_res_ATE}

\begin{figure}[htp!]
    \centering
    \includegraphics[width=0.7\linewidth]{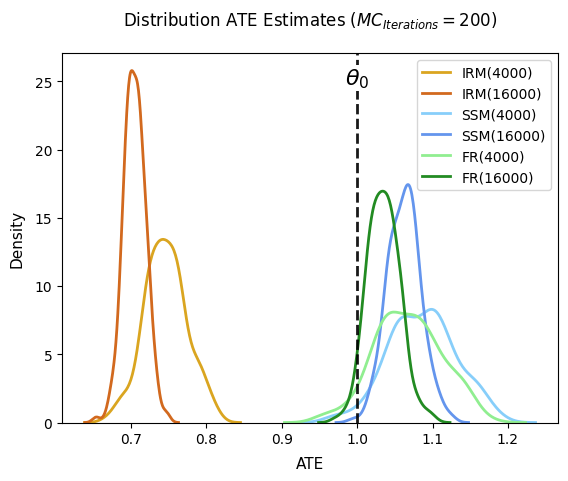} \\[-.43 cm]
    \caption{\normalfont This figure displays the distribution of ATE estimates based on $\theta_0 = 1$ and $200$ Monte Carlo iterations. It illustrates that as the sample size grows, the IRM suffers from an increasing downward bias, while both the SSM and FR converge to the simulated ATE.}
    \label{fig:simulation}
\end{figure}

\clearpage

\subsection{Additional Simulation Results: SSM}
\label{app_sub:sim_add_res_SSM}

\begin{table} [ht!]
\begin{center}
\begin{adjustbox}{max width=.88\textwidth}
\begin{threeparttable}
\begin{tabular}{c c ccc c ccc c ccc}
\hline \\[-2.5 ex]
&& \multicolumn{3}{c}{\textbf{Lasso/Logistic}} 
&& \multicolumn{3}{c}{\textbf{RandomForest}} 
\\[.07 ex]
\hline\\[-2.5 ex]
N && ATE & SE & MAE
&& ATE & SE & MAE \\[.07 ex]
\hline \\[-2.5 ex]
1000  
&& 1.0511 & 0.0460 & 0.0863
&& 1.1228 & 0.0450 & 0.1287  \\
4000 
&& 1.0254 & 0.0222 & 0.0393 
&& 1.0895 & 0.0222 & 0.0901 \\
16000
&& 1.0123 & 0.0111 & 0.0205 
&& 1.0653 & 0.0110 & 0.0653 \\[.16 ex]
\hline
\end{tabular}
\end{threeparttable}

%
\end{adjustbox}
\end{center}
\vspace{-.43 cm}
\caption{This table presents the average SSM simulation results based on $\theta_0 = 1$ and $200$ Monte Carlo replications. It demonstrates that when nuisance functions in the SSM are estimated using Lasso specifications, as in \cite{Bia2024}, the bias declines as expected,  highlighting the importance of proper hyperparameter tuning when applying random forest learners to estimate the true ATE within the SSM framework.}
\label{tab:ssm_only}
\end{table}

\hfill

\begin{figure}[ht!]
    \centering
    \includegraphics[width=0.7\linewidth]{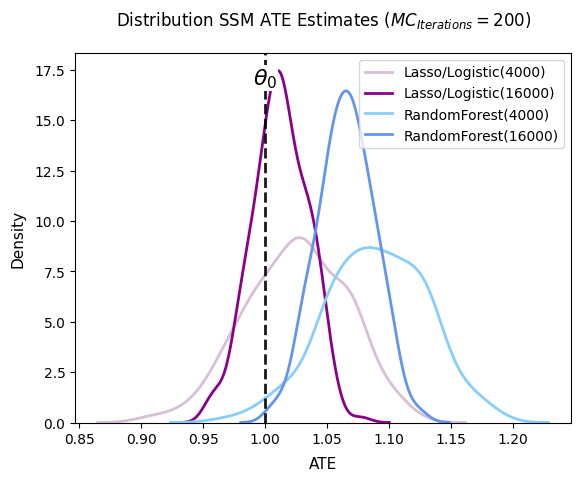} \\[-.43 cm]
    \caption{This figure displays the distribution of SSM ATE estimates based on $\theta_0 =1$ and $200$  Monte Carlo iterations. Given the linearity of the DGP, it illustrates that the Lasso/Logistic specification of \cite{Bia2024} converges faster to the true ATE than the random forest specification with $max\_depth = 20$. }
    \label{fig:ssm_only}
\end{figure}

\clearpage

\section{Additional Material for the Application}
\label{sec:tab_fig}

\begin{table}[!ht]
\begin{center}
\begin{adjustbox}{max width=\textwidth}
\begin{tabular}{l|c|c|l}
\toprule
\multicolumn{4}{l}{\textbf{Dependent variable: } S (reported wage indicator)} \\
\midrule
\hline \hline
\textbf{Interaction} & \textbf{Coef.} & \textbf{SE} & \textbf{p} \\
\midrule
\hline
Experience $\times$ Female & -0.0242 & 0.010 & 0.011** \\
Experience$^2$ $\times$ Female & 0.0007 & 0.0002 & 0.001*** \\
Household size $\times$ Female & 0.0993 & 0.021 & 0.000*** \\
Children $<$ 5 $\times$ Female & 0.0827 & 0.063 & 0.190 \\
Master degree $\times$ Female & -0.0173 & 0.058 & 0.764 \\
Professional degree $\times$ Female & 0.3203 & 0.071 & 0.000*** \\
Doctoral degree $\times$ Female & -0.1325 & 0.113 & 0.240 \\
Married (absent spouse) $\times$ Female & -0.0862 & 0.174 & 0.621 \\
Married (present spouse) $\times$ Female & -0.1838 & 0.072 & 0.011** \\
Never married $\times$ Female & 0.1465 & 0.085 & 0.083* \\
Separated $\times$ Female & -0.2638 & 0.217 & 0.225 \\
Widowed $\times$ Female & 0.1148 & 0.229 & 0.616 \\
Chinese $\times$ Female & -0.3191 & 0.189 & 0.091* \\
Other Asian $\times$ Female & -0.2551 & 0.132 & 0.053* \\
White $\times$ Female & -0.2321 & 0.091 & 0.011** \\
Not well English $\times$ Female & 0.2016 & 0.234 & 0.388 \\
English only $\times$ Female & 0.4320 & 0.147 & 0.003*** \\
English very well $\times$ Female & 0.4255 & 0.159 & 0.007*** \\
English well $\times$ Female & 0.4326 & 0.190 & 0.023** \\
Hispanic $\times$ Female & 0.0152 & 0.111 & 0.890 \\
Veteran $\times$ Female & -0.1159 & 0.180 & 0.519 \\
East South Central $\times$ Female & 0.3031 & 0.126 & 0.016** \\
Middle Atlantic $\times$ Female & 0.1689 & 0.086 & 0.048** \\
Mountain $\times$ Female & 0.1045 & 0.111 & 0.348 \\
New England $\times$ Female & -0.0324 & 0.104 & 0.756 \\
Pacific $\times$ Female & -0.0938 & 0.079 & 0.238 \\
South Atlantic $\times$ Female & -0.0907 & 0.082 & 0.270 \\
West North Central $\times$ Female & 0.1876 & 0.114 & 0.099* \\
West South Central $\times$ Female & 0.0895 & 0.093 & 0.335 \\
\bottomrule
\end{tabular} 
\end{adjustbox}
\end{center}
\caption{This table presents logit estimates for the probability of reporting  wages ($S=1$). Shown are only the interaction terms between female gender and key socio-economic characteristics. Positive coefficients indicate that the characteristic increases women’s reporting probability relative to men, whereas negative coefficients indicate the opposite. The results reveal substantial gender heterogeneity in wage reporting, suggesting that non-random selection into observed wages varies systematically across demographic groups.}
\label{tab:reporting}
\end{table}

\begin{table}[!ht]
\begin{center}
\begin{adjustbox}{max width=\textwidth}
\begin{tabular}{l|l}
\toprule
\multicolumn{2}{l}{\textbf{Dependent variable: } $\log(\text{wages})$} \\
\midrule
\hline \hline
\textbf{Variable} & \textbf{Interpretation} \\
\midrule
\hline
Age & Life-cycle earnings growth \\
Experience & Linear experience premium \\
Experience$^2$ & Concavity of returns to experience \\
College Degree & Returns to education \\
Married, spouse present & Household stability effect \\
Professional degree & Very high skill premium \\
Household size & Family composition \\
Never married & Labor supply differences \\
Pacific Division & Regional wage differences \\
Doctoral degree & Advanced education returns \\
\bottomrule
\end{tabular} 
\end{adjustbox}
\end{center}
\caption{This table reports the covariates most frequently used in the splitting rules of the ForestRiesz regression learner when predicting the outcome $Y$. Variables with higher split frequency are interpreted as having stronger predictive power for log(wages). The right column provides economic interpretations commonly associated with these predictors.} 
\label{tab:predictors}
\end{table}

\begin{table}[!ht]
\begin{center}
\begin{adjustbox}{max width=\textwidth}
\begin{tabular}{l|ccccccc}
\toprule
Group & $k$ &
$\theta_{\text{full}}$ &
$\theta_{-j}$ &
$\Delta\theta$ &
$|G_{Y,j}|$ &
$|G_{S,j}|$ &
$|\rho_j|$
\\
\midrule
\hline
Marital status & 5 & -0.1276 & -0.1331 & -0.00553 &  0.00177 &  0.00097 & 1.000 \\
Region         & 8 & -0.1276 & -0.1298 & -0.00225 &  0.00110 &  0.00174 & 1.000 \\
Race           & 3 & -0.1276 & -0.1287 & -0.00116 &  0.00115 &  0.00264 & 0.540 \\
Children       & 1 & -0.1276 & -0.1278 & -0.00027 &  0.00079 &  0.00138 & 0.210 \\
Education      & 3 & -0.1276 & -0.1275 &  0.00007 &  0.00435 &  0.01603 &  0.007 \\
Experience     & 2 & -0.1276 & -0.1275 &  0.00003 &  0.00003 &  0.00020 &  0.284 \\
\bottomrule
\end{tabular}

\end{adjustbox}
\end{center}
\caption{\normalfont Benchmarking results. Notes: $k$ = number of covariates dropped from group $j$. $\theta_{\text{full}}$ = ATE with full covariate set. $\theta_{-j}$ = ATE when covariate group $j$ is removed. $\Delta\theta = \theta_{-j} - \theta_{\text{full}}$ measures the sensitivity of the gender wage gap to group $j$. 
The sensitivity parameters $G_{Y,j}$, $G_{S,j}$, and $\rho_j$ are defined and explained in Appendix \ref{sec:benchmarking}.
Since the estimated sensitivity parameters reported in this table are approximately unbiased for the true shares, some of them can be negative when the true shares are close to zero.}
\label{tab:bench}
\end{table}

\begin{figure}
    \centering
    \includegraphics[width=0.8\linewidth]{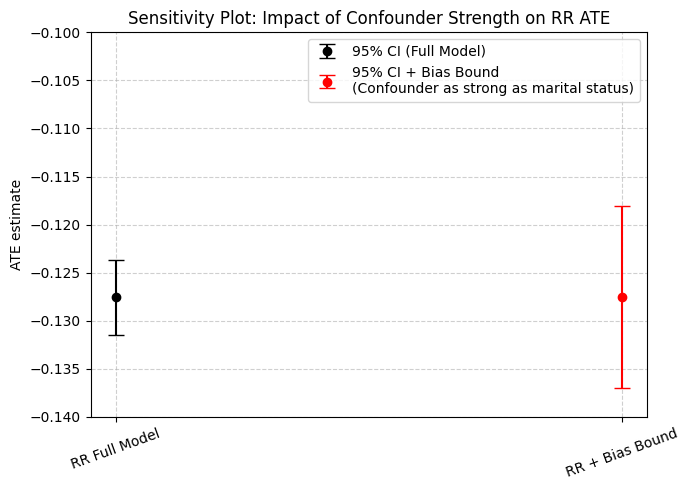}
    \caption{Sensitivity of the estimated gender wage gap (log wages) to potential omitted confounding. The plot compares the Riesz Representer ATE estimate from the full model with the counterfactual confidence interval that would arise if an unobserved confounder were as influential as marital status. While confidence intervals widen under this hypothetical confounder, the estimated ATE remains negative, indicating that the gender wage gap is robust to confounding of realistic magnitude.}
    \label{fig:CI}
\end{figure}

\begin{figure}
    \centering
    \includegraphics[width=1\linewidth]{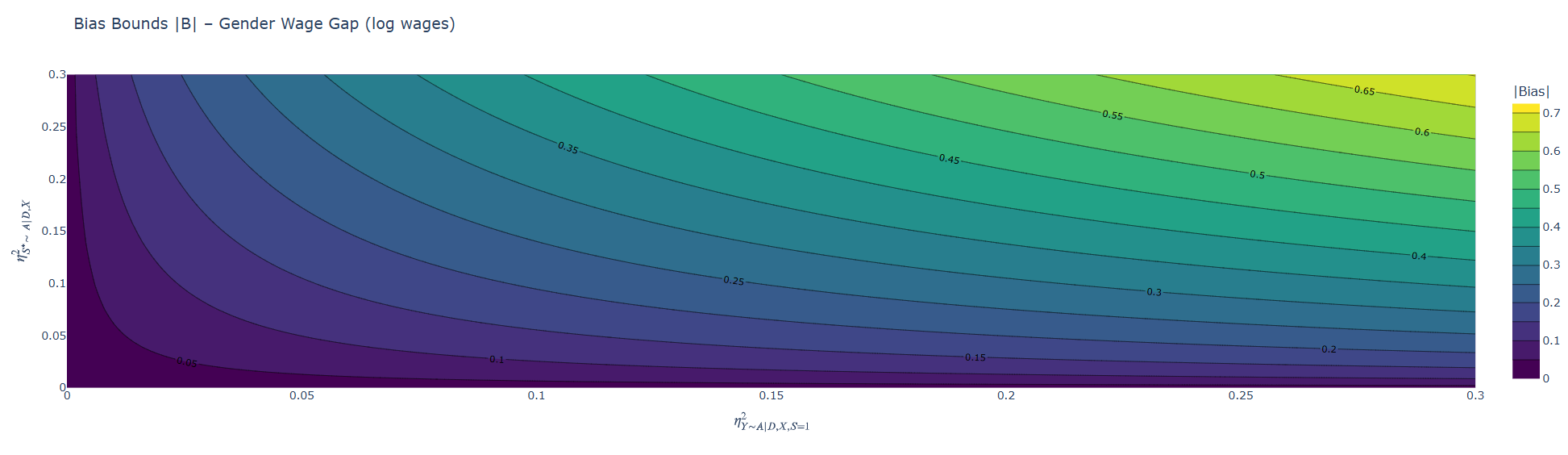}
    \caption{Contour plot of bias bounds as a function of outcome sensitivity $\eta^2_{Y \sim A | D,X ,S=1}$ and selection sensitivity $\eta^2_{S^*\sim~ A |D, X}$. The figure shows how large an omitted confounder must be in terms of explanatory power for both wages and selection into observed wages to overturn the observed gender wage gap. Only confounders with combined sensitivity above the robustness threshold ($RV = 0.063$) could eliminate the estimated effect, implying strong robustness to selection and outcome confounding.}
    \label{fig:sensitivity}
\end{figure}

\end{document}